\documentclass{aastex}
\usepackage{rotating}
\usepackage{epsfig}  
\usepackage{epstopdf} 
\usepackage{graphics}
\usepackage{listings}

\begin{document}


\title{The Detection of Discrete Cyclotron Emission Features in Phase-Resolved
Optical Spectroscopy of V1500 Cygni}

\author{Thomas E. Harrison}

\affil{Department of Astronomy, New Mexico State University, Box 30001, MSC 
4500, Las Cruces, NM 88003-8001}

\author{Ryan K. Campbell}

\affil{Department of Physics \& Astronomy, Humboldt State University, 1 Harpst S
t., Arcata, CA 95521}

\email{Ryan.Campbell@humboldt.edu}

\begin{abstract}
We have obtained phase-resolved optical spectroscopy of the old nova, and 
asynchronous polar V1500 Cyg. These new data reveal discrete cyclotron humps
from two different strength magnetic fields. One region has B = 72 MG,
while the other has B $\simeq$ 105 MG. With the detection of these features, we
revisit the optical/near-IR light curves presented in Harrison \& Campbell
(2016), and find that the large photometric excesses observed in those data are 
fully reconcilable with cyclotron emission. These results, when combined with 
the X-ray observations that appeared to have maxima that repeated on the 
orbital period, imply that V1500 Cyg has reverted back to a synchronous polar. 
Using existing theory, we show that the strong field strengths found here can 
explain the rapid spin down time.
\end{abstract}


\begin{flushleft}
$^{\rm 1}$Partially based on observations obtained with the Apache Point 
Observatory 3.5-meter telescope, which is owned and operated by the 
Astrophysical Research Consortium.\\
\end{flushleft}

\section{Introduction}

If there was a Pantheon for classical novae, V1500 Cygni (Nova Cyg 1975) would 
almost certainly occupy the central altar. It was one of the brightest, fastest
and most 
luminous classical novae (CNe) of all time. In the weeks before the eruption,
the progenitor brightened by $\sim$ 7 mag, unique among CNe (Collazzi et al.
2009). Stockman et al.  (1988) found that the light from V1500 Cyg was 
circularly polarized, indicating a highly magnetic system. It was a ``polar.'' 
Strangely, the period of the sinusoidal variation in the circular polarization 
was not synchronized to the orbital period, but to a period that was 2 per cent 
shorter. Unlike all other polars known at the time, the white dwarf in V1500 
Cyg was spinning with a period that was different from the orbital period. 
Apparently, the CNe eruption had knocked the system out of synchronism, and 
V1500 Cyg became the first identified ``asynchronous polar.'' Nearly as 
surprising was the finding that the white dwarf spin period was lengthening, 
and the system would again attain synchrony within $\sim$ 200 years (Schmidt 
\& Stockman 1991). 

To attempt to explain the rapid spin down time scale of V1500 Cyg and other
asynchronous polars, Campbell \& Schwope 
(1999) developed a model that used the torque resulting from induced 
electric fields and current flow in the secondary star from an asynchronously
spinning magnetic white dwarf to slow its rotation rate. In their modeling, if 
the white dwarf mass was of order 0.5 M$_{\odot}$, the spin-down time scale was 
of order 50 yr. If the white dwarf mass was higher than this, the spin-down 
time scale would be longer. Given the violence of the CNe eruption of V1500 
Cyg, and the presence of a modest enhancement of neon in its ejecta, it is 
assumed that the mass of the white dwarf in V1500 Cyg is large (Politano et 
al. 1995, Lance et al. 1988). Since the spin-down time scale is directly
proportional to the mass of the white dwarf, the time required to attain 
synchrony would be of order $\sim$ 100 yr. To agree with the observations of 
Schmidt \& Stockman (1991), simply requires the white dwarf to have a magnetic 
field strength that is of order B $\sim$ 28 MG.

Harrison \& Campbell (2016, hereafter ``HC16'') obtained $XMM-Newton$ 
observations of 
V1500 Cyg that showed a strong X-ray maximum once each orbit. Timing 
analysis of the five observed maxima in this light curve appeared to suggest 
that the white dwarf was now spinning at the orbital period---the system had 
become synchronized. This conclusion was supported by optical and infrared 
light curves that, while dominated by flux from the heavily irradiated 
secondary star, appeared to show excesses that always occurred at the same 
orbital phases. These data spanned seven years, and suggested that the system 
had attained synchrony as early as 2007. HC16 attributed the largest of these 
excesses to cyclotron emission from a pole with a magnetic field strength of 
$\approx$ 80 MG. This high magnetic field strength could explain the rapid 
spin down rate using the Campbell \& Schwope prescription. For a field strength 
near 80 MG, the $n$ = 2, 3, and 4 cyclotron harmonics would be present in 
optical spectroscopy. 

These results suggest that it should be feasible to detect discrete cyclotron 
harmonic emission in spectra of V1500 Cyg. We have obtained phase-resolved 
optical spectroscopy and find that the spectra of V1500 Cyg do exhibit 
cyclotron emission features throughout its orbit. We have attempted to model 
those features to estimate the parameters of the cyclotron emission regions. 
The phases at which the cyclotron emission occurs has lead to a 
complete revision of the light curve modeling presented in HC16. In the next 
section we discuss the data set, in section 3 we present the new phase-resolved 
spectra and model it, and the light curve data from HC16. We discuss these 
results in section 4, and present our conclusions in section 5.

\section{Data}

Observations of V1500 Cyg were obtained 2017 July 17 using the Double 
Imaging Spectrograph\footnote{http://www.apo.nmsu.edu/arc35m/Instruments/DIS/} (``DIS'') on the Apache Point Observatory 3.5 m. The two channels of
DIS allow simultaneous optical spectra in both the blue and the red, by using a 
dichroic mirror with a transition wavelength of 5350 \AA. This creates two
separate optical paths feeding separate gratings. The default low
resolution mode was used, delivering a dispersion of 1.8 \AA/pix in the
blue, and 2.29 \AA/pix in the red. The target was observed from 
07:17 to 10:51 UT, covering a full orbit (P$_{\rm orb}$ = 3.35 hr). The
exposure times for V1500 Cyg were all identical, 720 s. In the following,
we have used the minimum light ephemeris of Semeniuk et al. (1995) to
calculate the phases of our data set. As described below, the phasing
of this ephemeris does not correspond to inferior conjunction of the
secondary, but to the time of the $V$-band minimum. As shown in HC16,
the light curves in the $BVRIJHK$ bandpasses all have minima {\it near}
$\phi$ = 0.0. The maximum error in the Semeniuk et al. ephemeris, which
has a T0 date of 1986 September, at the epoch of our spectroscopic data set is 
$\Delta \phi$ = $\pm$ 0.13.

\begin{figure}[htb]
\centerline{{\includegraphics[width=15cm]{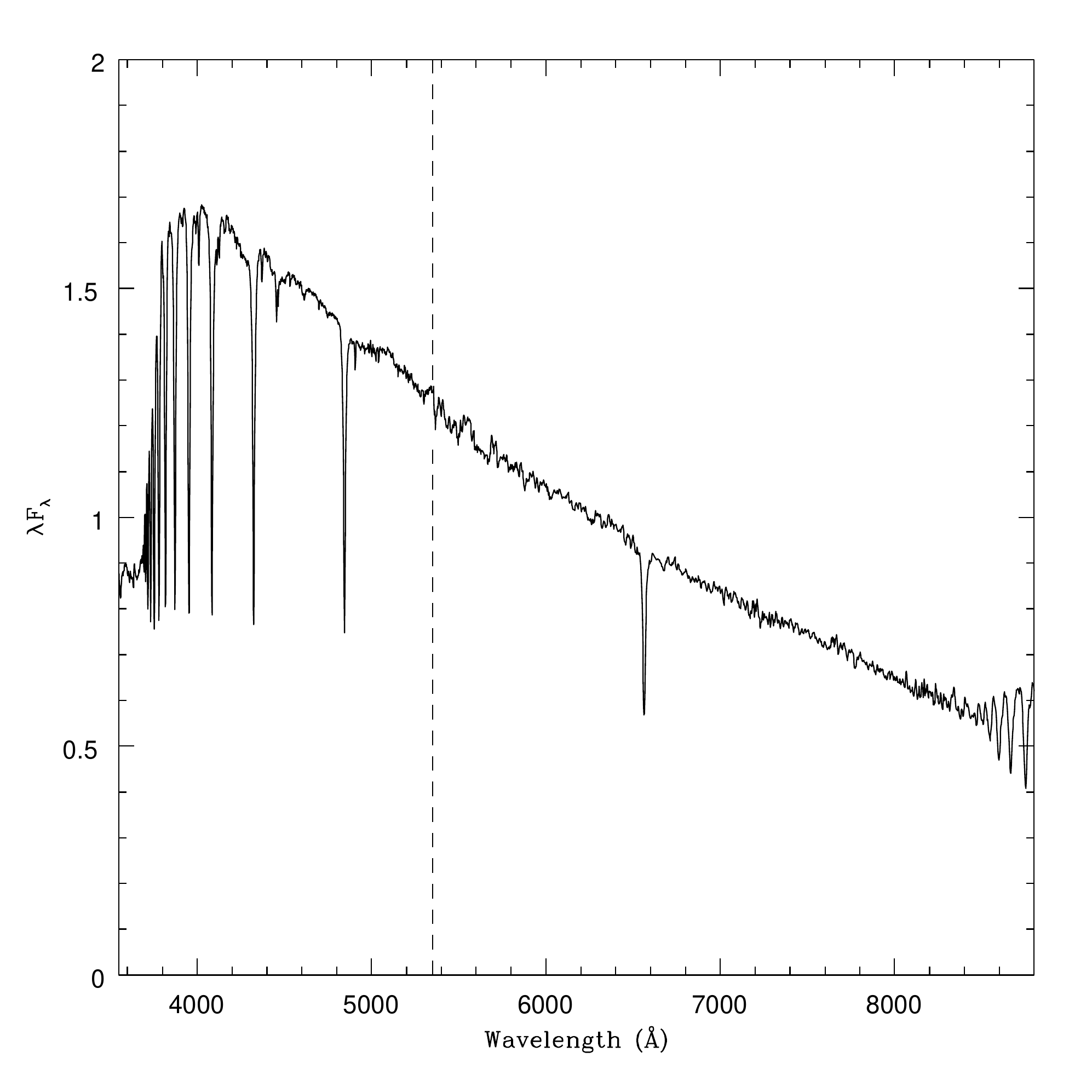}}}
\caption{The spectrum of HR 9087 after correction by the ``smooth star''
spectrum (see the text). The location of the transition wavelength for
the dichroic in the DIS spectrograph is indicated by the vertical, dashed
line.  }
\label{hr9087}
\end{figure}

The data were reduced in the usual fashion using {\it IRAF}. In addition, 
however, the techniques outlined by Bessell (1999), to use a ``smooth star'' 
to correct for telluric features in optical spectra, were also applied to
these data to aid in the identification of any weak cyclotron features.  
The white dwarf L1363-3 was observed immediately after finishing the 
observations
of V1500 Cyg. It was used to both remove the telluric features as well as 
to correct for any residuals that remained from the flux calibration process. 
As shown in Fig. \ref{hr9087} for one of our spectrophotometric calibrators
for the V1500 Cyg data set, HR 9087, this technique yielded excellent results,
with only small residuals in the Fraunhofer A band ($\lambda$7600 \AA). We also 
locate the dichroic transition 
wavelength in Fig. \ref{hr9087}. A residual exists at this wavelength in all 
of the DIS spectra, and becomes somewhat more substantial for fainter sources 
like V1500 Cyg. A small region around this location has been omitted
from the plots of the DIS spectra for V1500 Cyg. The raw spectra at $\phi$ = 
0.17 have S/N = 13 in the continuum at both 5000 \AA ~and 6000 \AA.

\section{Results}

The analysis of the $XMM$ data for V1500 Cyg in HC16 indicated that the spin 
period of the white dwarf was very close to that of the orbital period. 
Unfortunately, the X-ray maxima of V1500 Cyg are quite variable, and the $XMM$ 
observations were too short to confidently confirm synchronization. The 
compilation of optical/IR light curves presented in HC16 spanned some seven 
years. Over that time period the phasing and morphology of the light curves 
did not change. One could overlay the 2006 July 12 $JHK$ data on top of the 
30 October 2014 photometry, and they would be nearly indistinguishable. A 
similar result attains 
for the visual photometry, though V1500 Cyg exhibits larger variability at 
shorter wavelengths. It is not obvious how the excesses observed in the 
$BVRIJHK$ light curves could originate from the secondary star. Thus, they 
must come from a source associated with the white dwarf or the accretion 
stream. HC16 concluded that they were due to cyclotron emission from a 
synchronously rotating white dwarf.

The analysis of the X-ray data found two maxima. In the frame of the
Semeniuk et al. ephemeris, the stronger of these
occurred at $\phi$ = 0.82, with a secondary maximum occurring at $\phi$ = 0.32. 
HC16 modeled the X-ray light curve with two spots, with the primary accretion 
region on the white dwarf located below the line of sight. This meant that it 
was visible for less than one half of an orbital period. Due to its much lower 
luminosity, the location of the secondary accretion region was not well 
constrained. They simply placed it opposite to the primary region following
the results of Stockman et al. (1988), who concluded that the two accretion
regions responsible for the observed polarization were
on opposite sides of the white dwarf. Being in the upper hemisphere insures 
that the secondary pole remains visible for a longer time interval than the 
primary pole.

\subsection{The Orbital Evolution of the Spectra}

We present the phase-resolved spectra in four panels in Fig. 2. All of
the spectra for V1500 Cyg have been dereddened using the ``deredden'' task
in {\it IRAF} assuming A$_{\rm V}$ = 1.2. The deredden task 
employs the reddening law of Cardelli et al. (1989). Lance et al. (1988)
have compiled published values for the extinction to V1500 Cyg, and the
majority of them have A$_{\rm V}$ = 1.4. During the cyclotron modeling 
discussed in the next section (the coloured lines in Fig. 2), we found that 
an extinction of A$_{\rm V}$ = 1.4 mag produced a continuum that was too blue 
for us to match with any model that was comprised of a hot white dwarf, an
irradiated (cool) secondary star, and a cyclotron component. This
may be due to our na\"{i}ve assumptions about the underlying source, but models
matched to the data dereddened by A$_{\rm V}$ = 1.2 mag appeared to produce
sensible results with the minimum number of spectral components. 

The first panel, Fig. \ref{fig2a}, presents the spectra in the interval 0.17 
$\leq$ $\phi$ $\leq$ 0.36. A single broad hump, centred near 5400 \AA, is
present in each of the four spectra. As the system slowly brightens towards 
visual maximum ($\phi$ = 0.5), the symmetric hump grows to its largest size 
at $\phi$ = 0.23, and then declines to become much weaker by $\phi$ = 0.36. 
During the phase interval presented in Fig. \ref{fig2a}, the H I and He I 
emission lines are at their strongest, indicating that they might be partially
due to irradiation of the donor star. In the phase interval 0.43 $\leq$ $\phi$ 
$\leq$ 0.55, Fig. \ref{fig2b}, the hump remains relatively unchanged, though 
with a subtle shift in its peak to a longer wavelength at $\phi$ = 0.55. By 
$\phi$ = 0.62 the hump is very weak, and appears to be narrower. Over the 
interval $\phi$ = 0.68 to $\phi$ = 0.87, Fig. \ref{fig2c}, the continuum 
undergoes dramatic changes with the addition of humps centred near 
$\sim$ 4000 \AA ~and 7400 \AA. At $\phi$ = 0.0, Fig. \ref{fig2d}, near the time 
of the light curve minimum, the strongest hump seen in the preceding set of
spectra is now considerably smaller, is more symmetric, and appears to have 
shifted redward. The phase 0 spectrum also appears to have features that are 
suggestive of a late-type star. In Fig. \ref{M5comp} we plot the red side of 
the $\phi$ = 0.0 spectrum of V1500 Cyg, and compare it to that of the M5 dwarf 
Gl 866A (obtained using DIS). 

The rapid change in the morphology and strength of the broad hump over
an orbital period suggests that this feature can only be properly explained 
as due to cyclotron emission. As we describe below, we believe that there is 
actually cyclotron emission from two different regions with differing magnetic
field strengths. It just happens that the peak fluxes from their strongest 
cyclotron harmonics occur at {\it similar} wavelengths. Given that the
cyclotron emission features are strongest near $\phi$ = 0.85 suggests that at 
this time, the origin of most/all of the cyclotron emission is associated with 
the primary accreting pole. The more symmetric hump, seen from 0.17 $\leq$ 
$\phi$ $\leq$ 0.68, is then associated with the secondary pole. There appears 
to be cyclotron emission at all phases of the orbit.

\subsection{Cyclotron Modeling}

The appearance of a cyclotron emission spectrum 
depends on a variety of parameters. In the ``constant-$\Lambda$'' modeling 
prescription (see Schwope 1990), these are the magnetic field strength
(B), the temperature of the cyclotron emission region ($k$T), the viewing 
angle ($\theta$), and the log of the optical depth parameter ($\Lambda$). 
To construct cyclotron models, we employ the constant-$\Lambda$ code used by 
Campbell (2008), originally developed by Schwope (1990).
The X-ray observations modeled in HC16 provide us with information 
on the temperatures of the two poles, with weaker constraints on the locations 
of the emission regions. For the primary X-ray emission region, for
the two component model they used (absorbed blackbody and free-free sources) 
a temperature for the thermal bremsstrahlung component of $k$T = 40 $\pm$
9 keV was found. Their fit for the much fainter secondary pole found $k$T = 
1.85 $\pm$ 0.7 keV. Given the low flux, the properties of this pole were not 
tightly constrained. The duration 
of the primary X-ray maximum indicated that the pole associated with this 
emission was visible for less than one-half of an orbital period. HC16
located this pole at a co-latitude of $\beta_{\rm 1}$ = 120$^{\circ}$, placing 
it in the hemisphere that is below our line of sight. With the only constraint 
being the anti-phasing of the X-ray emission, the secondary pole was placed 
diametrically opposite to the primary pole giving $\beta_{\rm 2}$ = 
60$^{\circ}$. With a binary inclination of $i$ $\approx$ 60$^{\circ}$ 
(Schmidt et al. 1995), this region would in view for more
than one half of an orbit. For cyclotron modeling, such a co-latitude suggests 
that the viewing angle to the secondary pole changes very slowly over an orbit. 
The opposite is true for the primary pole, with larger viewing angles 
throughout its visibility. As discussed in HC16 (and references therein) low 
viewing angles to the emission region result in broader, less distinct
cyclotron harmonics. Higher viewing angles produce narrower cyclotron emission
features (for a much fuller discussion on how the parameters in a 
constant-$\Lambda$ model influence the appearance of cyclotron spectra, see 
Harrison \& Campbell 2015).

As shown in Campbell (2008), synthetic spectra produced from
constant-$\Lambda$ modeling generally
requires orbitally (phase) dependent values for B, $\Lambda$, and $k$T 
to match observations. Even if one has a relatively good 
understanding of the location of the cyclotron emission region, and of the 
underlying continuum source, the assumptions of a constant temperature 
accretion region, a single value of $\Lambda$, or one magnetic field strength, 
are never realized. It is also likely that the cyclotron emission region
is not a point source, and a single value for $\theta$ or B is unrealistic. It 
is also easy to obtain similar solutions by simultaneously changing the 
values of just two parameters. For example, B and $\theta$ are partially 
degenerate. Constant-$\Lambda$ models are also arbitrarily normalized. Thus, 
any solution that is derived from such modeling can only be assumed to be 
approximate, even when multiple harmonic features are observed. In the
modeling described below, the uncertainty about the orbitally evolving
continuum source, and our inability to constrain the location of the secondary 
pole, lead to additional difficulties in characterizing the cyclotron 
emission.

We start by modeling the spectra near $\phi$ = 0.25 (Fig. \ref{fig2a}), as
there is a single, symmetric hump at this time. HC16 modeled the $BVRIJHK$ 
light curves with two stellar components having temperatures of 52,000 K, 
and 3,100 K. The exact value of the temperature for the white dwarf is mostly 
irrelevant for the modeling here, since such a hot
source peaks far into the UV, and only the tail of this source's spectrum is
present in our data set. At superior conjunction, we estimate that the 
irradiated secondary has a temperature on the order of T$_{\rm eff}$ $\simeq$ 
6,100 K. Much of the rise to visual photometric maximum is due to the increased 
visibility of the donor's irradiated hemisphere. Thus, our models for the 
continuum of V1500 Cyg near $\phi$ = 0.25 comprise two blackbodies, the hot 
white dwarf primary, and the irradiated hemisphere of 
the secondary. We allowed the ratio of their contributions to be a variable at 
all times in this modeling. We discuss constraints on their relative 
luminosities in the next section. We also separately adjusted the strength 
of the cyclotron emission to best match each of the spectra.

We find that a cyclotron component with the following attributes explains
all four spectra in Fig. \ref{fig2a}: B = 105 MG, $k$T = 16 keV, $\theta$ =
60$^{\circ}$, and log$\Lambda$ = 0. The cyclotron hump centred near $\lambda$
= 5400 \AA ~is the $n$ = 2 harmonic for this field strength. The $n$ = 3
harmonic would be centred near 3600 \AA. The weakness of the
$n$ = 3 harmonic requires a low value for log$\Lambda$. As discussed above, 
these phases are associated with the visibility of the secondary accretion 
region. If it originates from a point-source, the broadness of 
the dominant cyclotron feature requires a temperature that is much higher than 
estimated from fitting its X-ray spectrum. We find that it is extremely 
difficult to reproduce the breadth of this feature with temperatures less than 
$k$T $\simeq$ 16 keV. A similar quality of fit can be easily achieved with 
slightly higher temperatures by simultaneously changing the viewing angle 
and the field strength. At much higher temperatures, $k$T
$>$ 20 keV, the shape of the $n$ = 2 harmonic becomes much more peaked on the 
blueward side, and does not resemble the observed feature. In Fig. \ref{fig2a} 
we present a deconvolution of the final model used to explain the $\phi$ = 0.29 
spectrum.

Because these spectra cover the time of transit of this region ($\phi$ = 
0.32), the viewing angle to this pole is near a minimum at this phase. If $i$ = 
60$^{\circ}$, and $\beta_{\rm 2}$ = 60$^{\circ}$, then $\theta$ = 0$^{\circ}$ 
at $\phi$ = 0.32. There should be very little detectable cyclotron emission at 
this time. The fact that it is still detectable suggests that $\beta_{\rm 2}$ 
$<$ 60$^{\circ}$, or that the cyclotron emission is offset from the magnetic 
pole. As discussed in Schmidt \& Stockman (1991), the broad sinusoidal
polarization maxima suggested large (or distributed) cyclotron emission 
regions. This could explain the limited change in the viewing angle found in 
the modeling above, as well as the fact that significant cyclotron emission
remains present during the transit of the magnetic pole. A distributed
cyclotron emission region would also lead to a broader harmonic from
the superposition of emission from regions with slightly different field
strengths, and viewing angles. This might help explain the much hotter 
temperature we derive here, versus the X-ray results. Note that the 
spectrum at $\phi$ = 0.17 appears to have a red excess 
above the model we have used to fit its cyclotron emission. We will find that
in addition to emission from the B = 105 MG pole, cyclotron emission from the 
primary pole, with its lower strength field, is necessary to explain several
of the spectra following the light curve minimum near $\phi$ = 0.0. 

For the next phase interval, 0.43 $\leq$ $\phi$ $\leq$ 0.62 (Fig. \ref{fig2b}), 
the cyclotron hump becomes weaker, and begins to change in shape. Given the
fact that we only have a single well-defined cyclotron harmonic,
we model the spectra at $\phi$ = 0.43 and 0.49 using the 
same parameters as used for the spectra in Fig. \ref{fig2b}. While it is 
subtle, the peak in the cyclotron emission appears to shift redward between 
$\phi$ = 0.49 and $\phi$ = 0.55. To fit the latter spectrum, we used same basic
model, but with a higher viewing angle: $\theta$ = 75$^{\circ}$. A changing 
viewing angle to the magnetic pole shifts the location of the wavelength of 
peak emission of a cyclotron harmonic (c.f., Schwope \& Beuermann 1997). 
Finally, to fit the very small, and poorly defined hump in the $\phi$ = 0.62 
spectrum, we used a model with the viewing angle increased to 85$^{\circ}$. 
Such a high viewing angle is consistent with the expectation that this
accreting pole is near the limb of the white dwarf at this phase. As we 
discuss in the next section, a small hotspot associated with the secondary 
pole would become self-eclipsed near $\phi$ = 0.71.

The dramatic change in morphology of the spectra over the phase interval
0.68 $\leq$ $\phi$ $\leq$ 0.87 (Fig. \ref{fig2c}) indicates that a different 
cyclotron emission spectrum has emerged, one associated with the primary
pole. We begin by modeling the
spectrum at $\phi$ = 0.87, as the distortions to the continuum are largest
at this time. We find that a cyclotron model with B = 72 MG, $\theta$ = 
62$^{\circ}$, $k$T = 35 keV, and log$\Lambda$ = 1.6 explains the data.
The broadness of the humps requires a very high temperature, in agreement
with the X-ray observations of the primary accretion region. The cyclotron
feature that peaks near 5200 \AA ~is the $n$ = 3 harmonic for this field 
strength. The blue
continuum shortward of 4800 \AA ~is well modeled by the $n$ = 4 harmonic
from this pole. The change in the slope of the continuum near 6500 \AA
~is fitted by the $n$ = 2 harmonic. At this phase, the cyclotron fundamental 
would peak near 1.35 $\mu$m. 

A nearly identical model, but with $\theta$ = 60$^{\circ}$, can be fitted to 
the $\phi$ = 0.81 spectrum. This spectrum coincides with the transit of the
primary pole, and the deconvolution of the model spectrum shown in Fig. 
\ref{fig2c} shows that it is comprised of just the combination of emission 
from a hot blackbody source (blue) and cyclotron emission (magenta). If the 
orbital inclination is 60$^{\circ}$, and $\beta_{\rm 1}$
= 120$^{\circ}$, the viewing angle should be $\theta$ = 60$^{\circ}$. The
presence of multiple harmonics in these spectra allows for more robust models 
to be constructed than was possible for the secondary pole. The $\phi$ =
0.75 spectrum is unusual. The expectation is that it should be dominated
by cyclotron emission from the primary pole. We could not find a model using 
only a cyclotron emission spectrum from the field with B = 72 MG that would 
reproduce 
this spectrum. The model (in red) is a composite of the B = 105 model
used to explain the spectra at earlier phases, and the B = 72 MG model used
at later phases. This is amplified by the fit to the $\phi$ = 0.68 spectrum,
where {\it only} the B = 105 MG ($\theta$ = 60$^{\circ}$, $k$T = 16 keV, and 
log$\Lambda$ = 0.0) model was used. 

Two of the spectra in Fig. \ref{fig2d} will require a similar two-component
cyclotron model. However, the spectrum at $\phi$ = 0.94 needs only emission
from the lower field strength region: B = 72 MG, $\theta$ = 50$^{\circ}$, 
$k$T = 35 keV, and log$\Lambda$ = 1.6. 
What is confusing is that if we have the lowest viewing angle to the emission 
region at $\phi$ = 0.94, the viewing angle to this pole at $\phi$ = 0.0 should 
probably be similar. As shown in Fig. \ref{fig2d}, we model the spectrum at
$\phi$ = 0.0 with a single component cyclotron model, but with that from the 
B = 105 MG field! After $\phi$ = 0.0, we see the return of emission from the 
cyclotron component with B = 72 MG, but now in combination with emission from 
the B = 105 MG pole. It is clear that emission from the B = 72 MG field is 
dominant in the spectra for $\phi$ = 0.06 and 0.11, but the broadness of the 
peak at 5400 \AA ~suggests that emission from the B = 105 MG pole is also
contributing to the spectra at these phases. These results argue that the 
cyclotron emission from the lower strength field is not co-located with 
the primary X-ray pole. 

\subsection{Reconciliation of the Cyclotron Emission with the $UBVRI$ Light 
Curves}

In HC16 the standard interpretation for the $V$-band light curve of V1500 Cyg
was employed, where the visual maximum that occurred at phase 0.5 was due
solely to the irradiated secondary star. With this phasing, $\phi$ = 0 would 
correspond to the inferior conjunction of the donor. If true, this 
leads to a large excess in the $BVRIJHK$ light curves in the interval 
0.0 $\leq$ $\phi$ $\leq$ 0.4 (see Fig. 5 in HC16). HC16 found that they
could not reconcile this large excess with cyclotron emission from either of 
the two X-ray poles without dramatically offsetting its location well away 
from the poles. Given that there is strong cyclotron emission near the 
time of primary X-ray pole transit, indicates that the absolute phasing of 
the underlying binary used in HC16 (and presumed elsewhere) is wrong.

Insight into the true phasing of the stellar components in V1500 Cyg
can be drawn from the $JHK$ light curves in HC16 (their Fig. 7). The maximum 
in those light curves 
occurs at $\phi$ = 0.32, exactly at the phase where the primary X-ray pole 
would be facing away from the observer. Given the high temperature of this 
source, it would help increase the irradiation of the donor, leading to a light 
curve maximum. As we have just shown, there is cyclotron emission at all
phases, thus the exact phasing of the secondary in its orbit remains
somewhat uncertain. Looking at the $K$-band light curve, perhaps the step 
stretching from $\phi$ = 0.4 to $\phi$ 0.5 in these data actually
arises from the irradiated secondary, and is not due to cyclotron emission.
If so, it is possible that the secondary could lag the primary X-ray pole by 
as much as 0.10 in phase. We find that we are able to construct a more robust 
set of light curve models with the assumption that inferior conjunction of the 
secondary star occurs near $\phi \simeq$ 0.82. This change in perspective 
shifts the period of the largest observed excesses in the optical/IR light 
curves to the phase interval 0.7 $\leq$ $\phi$ $<$ 1.0, associating them with 
the visibility of the primary pole, and the period when the cyclotron features
in the optical spectra are at their strongest.  

Before we attempt to construct new models for the $UBVRIJHK$ light curves
of V1500 Cyg, we attempt to characterize the stellar components in the system.
The light curves in HC16 (their Fig. 6) have $V$ = 19.35 at minimum light. The 
distances to CNe are difficult to estimate. Lance et al. adopted $d$ = 
1.2 kpc for V1500 Cyg from a variety of published values, while Slavin et al. 
(1995) found $d$ = 1.5 kpc from a nebular expansion parallax. If we use $d$ 
= 1.2 kpc, we derive M$_{\rm V}$ = 7.8 at minimum light. Given that cyclotron 
emission is always present, this is the upper limit to the 
combined luminosity of the primary and secondary stars (plus any contribution
from a hotspot). A normal M4V has M$_{\rm V}$ = 12.71 (Bessell 1991), while 
the Roche-lobe filling secondary star of U Gem (spectrally classified as an 
M4, but see Harrison 2016) has M$_{\rm V}$ $\simeq$ 10.1. At minimum 
light the luminosity of V1500 Cyg is dominated by the white dwarf primary,
since we are (mostly) viewing the un-irradiated hemisphere of the secondary.
White dwarfs that have absolute visual magnitudes near that of V1500 Cyg all 
have high temperatures, T$_{\rm eff}$ $\sim$ 40,000 K (Vennes et al.  1997). 
Though nearly all of those white dwarfs have lower masses (M$_{\rm wd}$ $\leq$ 
0.7 M$_{\sun}$) than that expected for V1500 Cyg (M$_{\rm 1}$ $\gtrsim$ 1 
M$_{\sun}$, see HC16).

At the time of our best view of the irradiated hemisphere of the secondary star,
$\phi \simeq$ 0.32, V1500 Cyg has $V$ = 18.4, or M$_{\rm V}$ = 6.8. Assuming
the white dwarf has the same $V$-band luminosity at all phases, implies that 
the irradiated secondary star has M$_{\rm V}$ = 7.3. The two stellar components 
supply somewhat similar fluxes at this phase. With an estimated radius of 0.42 
R$_{\sun}$, the irradiated hemisphere of the donor star has T$_{\rm eff} 
\sim$ 5200 K. The observed colour at this time is ($B - V$) = 0.4, and with 
A$_{\rm V}$ = 1.2, ($B - V$)$_{\rm 0}$ = 0.0. To arrive at this combined
colour suggests that the white dwarf itself must have ($B - V$)$_{\rm 0} 
\lesssim$ $-$0.4, corresponding to a temperature in excess of 40,000 K 
(Cheselka et al. 1993). Again, there is cyclotron emission in the $V$ bandpass 
at this phase, making such estimates highly uncertain. It appears that there 
is little cyclotron emission in the infrared at phases near maximum. At this 
time ($J - K$) = 0.5, and with A$_{\rm V}$ = 1.2, is consistent with an object 
that has T$_{\rm eff}$ = 6100 K.

In Fig. \ref{optlc} we re-plot the $UBVRI$ light curves of V1500 Cyg from 
HC16.  Given the excellent observing conditions, we have extracted broadband 
magnitudes from the new spectra to create light curves. Given the low S/N of 
the bluest portion of the spectra, the bandpass used to generate the
``$U$-band'' light curve only spans the wavelength interval from $\lambda$ =
3656 \AA ~to $\lambda$ = 3826 \AA ~(versus $\lambda$3656 $\pm$ 170 \AA ~for 
Johnson $U$). Due to the dichroic cutoff, the $V$ bandpass, nominally 
5450 $\pm$ 500 \AA, had to be truncated at $\lambda$ = 5200 \AA. Thus, the 
light curves in these two bands only approximate the true $U$- and $V$-band 
fluxes. To normalize the data set in each band, we offset each of the light 
curves so that the flux that falls closest to that of the model light curves, 
falls exactly on the light curves. For the $U$-band this occurred at $\phi$ = 
0.87.  For all of the other bandpasses, this occurred at $\phi$ = 0.36, near 
the time of the transit of the secondary pole. The similarity between the 
spectroscopic light curves and the photometric light curves is striking, 
while also highlighting changes in the apparent strength of the 
cyclotron emission.

We locate the phases of the pole transits with vertical lines (solid for the
primary maximum, and dashed for the secondary maximum) in Fig. \ref{optlc}. 
Presuming a synchronous system, we use the Wilson-Devinney code 
(WD2010\footnote{ftp.astro.ufl.edu/pub/wilson/lcdc2010/}) to produce models, 
but now require the phase of the irradiation-induced maximum to occur at $\phi$
 = 0.32. For the model presented in Fig. \ref{optlc}, we assume a secondary 
star with T$_{\rm eff}$ = 3000 K, and a binary inclination of $i$ = 
60$^{\circ}$. We then let the temperature of the white dwarf be a free 
parameter. In addition, we add a hotspot (with a size and temperature to be 
determined from the modeling) on the white dwarf located at the 
longitude that corresponds to the primary X-ray maximum. HC16 found that the 
primary X-ray region appears to be visible for about $\Delta \phi$ = 0.18. As 
in that paper, we assume that the duration of this transit is due to the 
``southerly'' co-latitude of the accretion region. We find that $\beta_{\rm 1}$
= 128$^{\circ}$, versus $\beta_{\rm 1}$ = 120$^{\circ}$ derived in HC16, is 
more consistent with the light curve data.

The most stringent requirement for modeling the light curves of V1500 Cyg is to 
reproduce the large amplitude variations seen in the near-infrared. 
This amplitude is almost completely due to the irradiation
of the donor star, and thus is a sensitive measure of the temperature of the 
two stellar components in the system. To get a significant temperature change
on the irradiated hemisphere of the donor requires it to have a cool 
temperature. Secondaries with T$_{\rm eff}$ $\geq$ 4,000 K cannot produce
the observed variations. Given the relatively short orbital period of V1500
Cyg, and the results in Fig.  \ref{M5comp}, we have chosen to fix the donor 
star temperature to T$_{\rm eff}$ = 3000 K. Hotter donor star temperatures
will require hotter white dwarf temperatures to get the same response, 
while a cooler secondary star can be irradiated with a cooler white 
dwarf to recreate the observed variations.

We are able to explain the large amplitude seen in the near-infrared with two 
different models. The first has T$_{\rm wd}$ = 65,000 K, and the second has 
T$_{\rm wd}$ = 54,000 K. The difference is the value of the bolometric 
reflection albedo (``$w$'') used for the secondary star. As discussed in HC16, 
normally, for late-type 
convective stars, the bolometric reflection albedo is $w$ = 0.5. However, 
Barman et al. (2004) found that the intense irradiation of a cool star by a 
hot white dwarf should result in a dramatic change to its bolometric 
reflection albedo such that it should be like those of radiative 
stars ($w$ $\simeq$ 1). Barman et al. found that for a situation similar to 
that for V1500 Cyg, the bolometric albedo should be closer to $w$ = 0.8. With 
this value of $w$, the large amplitude of the $JHK$ light curves can be 
reproduced with the lower temperature white dwarf. Note that we can also
increase the light curve amplitude by increasing the orbital inclination.
This would reduce the required white dwarf temperature in either scenario.
However, we have started at $i$ = 60 $^{\circ}$, and thus have little room
to increase the inclination before eclipses begin at $i$ $\geq$ 72$^{\circ}$.
Models with $w$ = 0.5, $i$ = 71$^{\circ}$, and T$_{\rm wd}$ = 60,000 K,
can reproduce the $BVRIJHK$ data set.

The final model light curves presented in Figs. \ref{optlc} and \ref{irlc} 
have the temperature of the white dwarf as T$_{\rm eff}$ = 54,000 K, and the 
temperature of the hotspot associated with the primary pole of
T$_{\rm eff}$ = 432,000 K. A radius for the hotspot of R$_{\rm 1}$ = 
28$^{\circ}$ was found to best fit the $B$-band light curve. The WD2010 code 
will not allow for a much hotter temperature spot than this (the limit is 
500,000 K), so to get significant optical emission and irradiation requires a 
sizable spot radius. The resulting light curve models explain the rise to 
maximum in the {\it BVRI} light curves, but leave significant excesses that 
peak near $\phi$ = 0.5. This model almost fully explains 
the $JHK$ light curves (Fig. \ref{irlc}) from $\phi$ = 0.9 to $\phi$ = 0.4. 

If we again assume that the secondary accretion region is truly located 
exactly opposite the primary pole, it would have $\beta_{\rm 2}$ = 
51$^{\circ}$. We do not have a hotspot at this location in the 
model, as it appeared to contribute little to the X-ray light curve. If the
spot is geometrically small, with a radius of $\sim$ 3$^{\circ}$, it would
be visible for nearly 80 per cent of the orbit. Obviously, if the hotspot 
associated
with the secondary accretion region was as large as the primary accretion
region, some portion of the secondary hotspot would always be visible. As
noted earlier, the southerly latitude of the primary pole means that it is 
only visible for a smaller part of
the orbit. We can delineate the transit duration of each of the two accretion 
regions in the light curve plots (magenta and cyan arrows in the $B$-band 
panel, and corresponding dotted vertical lines) assuming these spot locations. 
For Figs. \ref{optlc} and \ref{irlc}, we have assumed the hotspot radius 
associated with the secondary pole is R$_{\rm 2}$ = 3$^{\circ}$, and that
for the primary is R$_{\rm 1}$ = 28$^{\circ}$.

Due to their simplicity, we examine the new models for the $JHK$ 
light curves first. It is apparent that the largest excess in these light
curves is associated with the visibility of the primary pole.
There also appears to be evidence for smaller excesses from $\phi$ = 0.95 to 
$\phi$ = 0.32, and from $\phi$ = 0.5 to $\phi$ = 0.7. Those excesses
must be associated with the secondary accretion region. The drop in the
flux in each of the three bandpasses after 
$\phi$ = 0.7, suggests that the cyclotron emission at this time is rapidly 
declining. The optical spectra, however, show that there is significant
cyclotron emission from the primary pole at $\phi$ = 0.94. Given the inherent 
variability of cyclotron emission, as demonstrated in the two sets of $JHK$ 
light curves in HC16 (their Figs. 7 and 8), perhaps the accretion rate simply 
declined as the system approached $\phi$ = 1.0 at the epoch of the infrared
observations. The more likely explanation, however, is that there is cyclotron 
emission from the secondary pole throughout most of the orbit, and the model 
light curves need to be shifted downwards by $\gtrsim$ 0.2 mag. This conclusion
is directly supported by the models to the spectra that show that cyclotron 
emission from the 105 MG field is present from $\phi$ = 0.0, to $\phi$ = 0.75. 

For the two fields used to model the cyclotron emission above, any 
near-infrared emission must come from the cyclotron fundamentals. The cyclotron 
fundamental is very broad at the high temperatures and values of log$\Lambda$ 
found in the two sets of models. For the 72 MG field model at $\phi$ = 0.87,
the $n$ = 1 harmonic has its peak flux near 1.35 $\mu$m. For the 105 MG
field, the cyclotron fundamental peaks near 9500 \AA. As discussed in 
Harrison \& Campbell (2015), unlike the higher harmonics, the cyclotron
fundamental has its peak emission at low viewing angles (see their Fig. 6).
It is interesting that all three light curves show a small peak right at
the time of transit of the secondary pole. There is also an inflection
in the $H$ and $K$-band light curves at the time of the primary pole transit.
The photometric excesses observed in the near-IR appear to be completely 
consistent with a cyclotron interpretation.

Comparison of the models to the optical light curves, shows that the 
largest excesses occur in the phase interval $\phi$ = 0.32 to $\phi$ = 1.0. 
For the photometry shown in Fig. \ref{optlc} for 31 August 2014, the peak in 
the $BVRI$ light curves occurred at a time when both poles were visible. A 
similar result is found from the spectroscopically-derived light curves. 
Except for the $V$-band light curve, the spectroscopic photometry is very
similar to the photometric light curve. The deviation of the $V$-band is 
easily explained due to the fact that the truncated bandpass only captures a 
fraction of the flux from the dominant harmonics of either field.

The light curve model for the $B$-band clearly shows the presence of the 
hotspot region associated with the primary X-ray pole. Due to its extreme 
temperature, it has little influence on the data beyond the $V$-band. It is 
guaranteed that much of the excess that we have wholly attributed to the 
hotspot in the $B$-band, arises from cyclotron emission from the primary pole, 
and that this hotspot is actually smaller than we have assumed here. The
spectroscopic light curves in $V$, $R$, and $I$ all show a small dip at 
$\phi$ = 0.82, relative to the photometric data. In contrast, the spectroscopic
$BVRI$ light curves from phase 0 to phase 0.2 show an excess above the 
photometric light curves. This type of variability is common among magnetic
CVs, but the constancy of the phasing of the light curves, and their excesses,
is further support for a synchronized system.

The $U$-band light curve that was difficult to understand using the old
binary phasing, now appears to be more comprehensible. As discussed in HC16, 
there are times when the flux in this bandpass suddenly drops to very low 
levels. These minima allow us to set the floor for the two-component light 
curve model. The maximum in the $U$-band photometry now occurs close to the 
time of the transit of the primary X-ray emission region, with its associated 
hotspot. While some of the excess emission in this bandpass is probably due 
the H I Balmer continuum, it is important to stress that the $n$ = 3 harmonic 
for the 105 MG field peaks near 3500 \AA. Unfortunately, our spectra do not go 
far enough blueward to fully appreciate its contribution. This harmonic, 
however, supplies much of the flux shortward of 4500 \AA ~for the spectra in 
Fig. \ref{fig2b}. Given this fact, and the presence of the $n$ = 4 harmonic of 
the B = 72 MG field in the $U$-band, the strong excesses above the light curve 
model throughout the orbit can be directly attributed to cyclotron emission.

While the spectroscopic $U$-band light curve curve is consistent with the
data from XMM for most of the orbit, right near the time of the transit
of the primary pole, there was a sudden drop in flux. This might be due to
the red bias of the synthetic $U$-band. This dip, however, is mirrored in the 
$VRI$ light curves. Perhaps the cyclotron emission at the epoch of the
spectroscopy was weaker than it was at the time of the photometry. Given that 
the cyclotron emission from the primary pole was still present in the spectra 
at these phases, provides additional support for a small offset of the model 
light curves to fainter magnitudes.

\section{Discussion}

The analysis of the $XMM$ observations in HC16 were surprising, suggesting
that the asynchronous polar V1500 Cyg had attained synchrony much earlier
than predicted. It had reverted to being a synchronous polar. 
The optical and infrared light curves obtained at various epochs over the past 
decade also supported the interpretation of a synchronous 
system: the main excesses/distortions in these light curves repeated at the 
same phases. If V1500 Cyg was now a synchronous polar, the strong light curve
excesses suggested that emission from discrete cyclotron humps {\it could}
be visible. We have obtained phase-resolved spectra of V1500 Cyg that show
such features, and they change in size and shape over the orbit like the
cyclotron features seen in other polars. The $BVRIJHK$ light curve analysis 
in HC16 estimated field strengths for the two cyclotron emission regions of 
$\sim$ 80 MG and $\geq$ 120 MG. Our new modeling of the spectra finds 
B$_{\rm 1}$ = 72 MG, and B$_{\rm 2}$  $\simeq$ 105 MG. The breadth
of the cyclotron humps from the primary pole leads to a temperature that is
consistent with that derived from the $XMM$ data set. We find, however, that
to explain the profile of the cyclotron humps from the secondary pole
requires a much hotter temperature than indicated by the X-ray observations.
Unfortunately, the complexity of the light curve from the underlying binary, 
combined with the coarse phase resolution and modest S/N of our data set, limit
our ability to accurately constrain the properties of the cyclotron emission 
regions. This is especially true for the secondary pole. To achieve more
robust results for {\it both} poles will require better constraints on the
cyclotron emission from the secondary pole, since emission from
this pole appears to be present at some level for nearly the entire orbit.

HC16 found it to be extremely difficult to attribute the excesses in
the light curves of V1500 Cyg with cyclotron emission if the phase of the
$V$-band maximum at $\phi$ = 0.5 was superior conjunction of the donor. They 
required large offsets of the cyclotron emission regions from the X-ray poles, 
and bent field lines, to produce the observed modulations. If we shift the
time of superior conjunction to correspond to the transit of the secondary
pole, $\phi$ = 0.32, the photometric excesses are easily explained.
The largest excesses in the visual data occur during phases when both 
poles are visible. 

While the new phasing of the light curves leads to photometric excesses
that are now consistent with the locations of the two X-ray poles, this is not 
as true for the spectra. The main issues occur near phase $\phi$ = 0.7, and 
between phases 0.94 and 0.17. Given our limited ability to constrain the 
location of the secondary pole, the period of its visibility remains difficult 
to quantify. Using a very small spot radius for this pole suggests that it
would become self-eclipsed near $\phi$ = 0.71. It appears that
this event actually occurs closer to $\phi$ = 0.62. However, by $\phi$ = 0.68, 
cyclotron emission that we associated with this pole is again present! It 
does not fully disappear until after $\phi$ = 0.75. One simple explanation
for this behavior is that it is not located exactly opposite to the primary
pole. Comparison of the spectroscopically derived light curves to the
photometric light curves during this phase interval shows that they are quite
similar, and thus the change in the morphology of the spectra seen at these
phases does not appear to be atypical. 

The emergence of the cyclotron humps from the primary pole near $\phi$ = 0.75 
is later than expected given the phasing of the X-ray pole. The self-eclipse
of this region should occur near $\phi$ = 0.0. While the cyclotron emission
from this pole {\it does} disappear at this phase, it returns at $\phi$ = 0.06,
and appears to remain visible up to $\phi$ $\simeq$ 0.17. The only explanation
we can come up with is that the cyclotron emission region associated with the 
primary pole is offset by about $+$0.15 in phase. This localization
would mean that it rises and sets later than predicted. The disappearance
of the cyclotron emission from this pole at $\phi$ = 0.0 might be attributable
to the transit of the emission region across our line of sight. Rapid declines 
in the cyclotron emission can occur when we look down the accretion 
funnel (see Bailey et al. 1982). In this geometry, we are looking along
the accretion stream, and this material can obscure our line of sight
to the cyclotron emission region. The H I emission lines are at their 
weakest near $\phi$ = 0, which is also consistent with such an event (see 
Howell et al.  2008). Additionally, the low values for log$\Lambda$ that we
found in modeling the emission from the primary pole also suggests that this
region is offset from the X-ray pole. Given the large accretion rate, and 
high temperatures implied by the X-ray data, one would expect the main
shock region to have very high optical depth. This was not found in our
modeling.

\section{Conclusions}

Given that 1) we know of no asynchronous polars whose spectra show discrete
cyclotron harmonic features, 2) we have an X-ray data set that suggests a spin 
period
similar to the orbital period, and 3) a relatively consistent photometric 
behavior that can be fully explained by the observed cyclotron emission, we 
conclude that V1500 Cyg has again become a synchronous polar, well ahead
of expectations. As noted in the introduction, Campbell \& Schwope (1999) 
developed a model that relied on a torque on the white dwarf that resulted 
from the currents and electric fields induced in the secondary by the 
asynchronous rotation of the magnetic primary to estimate the spin down 
time-scale of the white dwarf in V1500 Cyg. Our data set
suggests that synchrony occurred within $\sim$ 32 years of the CNe 
eruption. Using the Campbell \& Schwope formulation, if the white dwarf mass 
in V1500 Cyg is 1.0 M$_{\sun}$, and the magnetic field strength is 40 MG, 
the spin down time would be 110 yr. However, the synchronization time-scale is 
dependent on B$^{\rm -2}$. Using B = 72 MG, the time-scale for synchronization 
of the white dwarf in V1500 Cyg is only 34 yrs. With all of the uncertainties 
that go into this estimate, the result is remarkable.

To fully understand V1500 Cyg will require additional epochs of spectroscopy. 
Data obtained on a larger telescope are
necessary to both improve the S/N ratio, and provide better phase
resolution. There might also be epochs when the cyclotron emission from the 
secondary accretion region is much stronger than what we observed. If so, it 
might be possible to develop more robust cyclotron models for this pole. It 
would also be interesting to obtain new phase-resolved near-IR spectroscopy to 
search for emission from the cyclotron fundamentals. We re-examined the $JHK$ 
spectra presented in Harrison et al. (2013, obtained with NIRC on Keck in 
2006), but could not find any evidence to support the presence of discrete 
cyclotron harmonic features in those data. 

\acknowledgements Based on observations obtained with the Apache Point 
Observatory 3.5-meter telescope, which is owned and operated by the 
Astrophysical Research Consortium. We thank the anonymous referee for 
useful comments.

\begin{center}
{\bf References}
\end{center}
Bailey, J., Hough, J. H., Axon, D. J., Gatley, I., Lee, T. J., et al. 1982,
MNRAS, 199, 801\\
Barman, T. S., Hauschildt, P. H., \& Allard, F. 2004, ApJ, 614, 338\\
Bessell, M. S. 1999, PASP, 111, 1426\\
Bessell, M. S. 1991, AJ, 101, 662\\
Campbell, C. G. \& Schwope, A. D. 1999, A\&A, 343, 132\\
Campbell, R. K. 2008, PhD thesis, New Mexico State Univ.\\
Cardelli, J. A., Clayton, G. C., \& Mathis, J. S. 1989, ApJ, 345, 245\\
Cheselka, M., Holberg, J. B., Watkins, R., \& Collins, J. 1993, AJ, 106, 2365\\
Collazzi, A. C., Schaefer, B. E., Xiao, L., Pagnota, A., et al. 2009, AJ, 138,
1846\\
Harrison, T. E., \& Campbell, R. D. 2016 (HC16), MNRAS, 459, 4161\\
Harrison, T. E., \& Campbell, R. D. 2015, ApJS, 219, 32\\
Harrison, T. E., Campbell, R. D., \& Lyke, J. E. 2013, AJ, 146, 37\\
Howell, S. B., Harrison, T. E., Szkody, P., Walter, F. M., \& Harbeck, D. 2008.
AJ, 136, 2541\\
Lance, C. M. McCall, M. L., \& Uomoto, A. K. 1988, ApJS, 66, 151\\
Politano, M., Starrfield, S., Truran, J. W., Weiss, A., et al. 1995, ApJ, 448,
807\\
Schmidt, G. D., Liebert, J., \& Stockman, H. S. 1995, ApJ, 441, 414\\
Schmidt, G. D., \& Stockman, H. S. 1991, ApJ, 371, 749\\
Schwope, A. D., \& Beuermann, K. 1997, AN, 318, 2\\
Schwope, A. D. 1990, RvMA, 3, 44\\
Semeniuk, I., Olech, A., \& Nalezyty, M. 1995, ActaAst, 45, 747\\
Slavin, A. J., O'Brien, T. J., Dunlop, J. S. 1995, MNRAS, 276, 353\\
Stockman, H. S., Schmidt, G. D., \& Lamb, D. Q. 1988, ApJ, 332, 282\\
Vennes, S., Thejll, P. A., Genova Galvan, R., \& Dupuis, J. 1997, ApJ, 480,
714\\

\clearpage
\renewcommand{\thefigure}{2a}
\begin{figure}[htb]
\centerline{{\includegraphics[width=15cm]{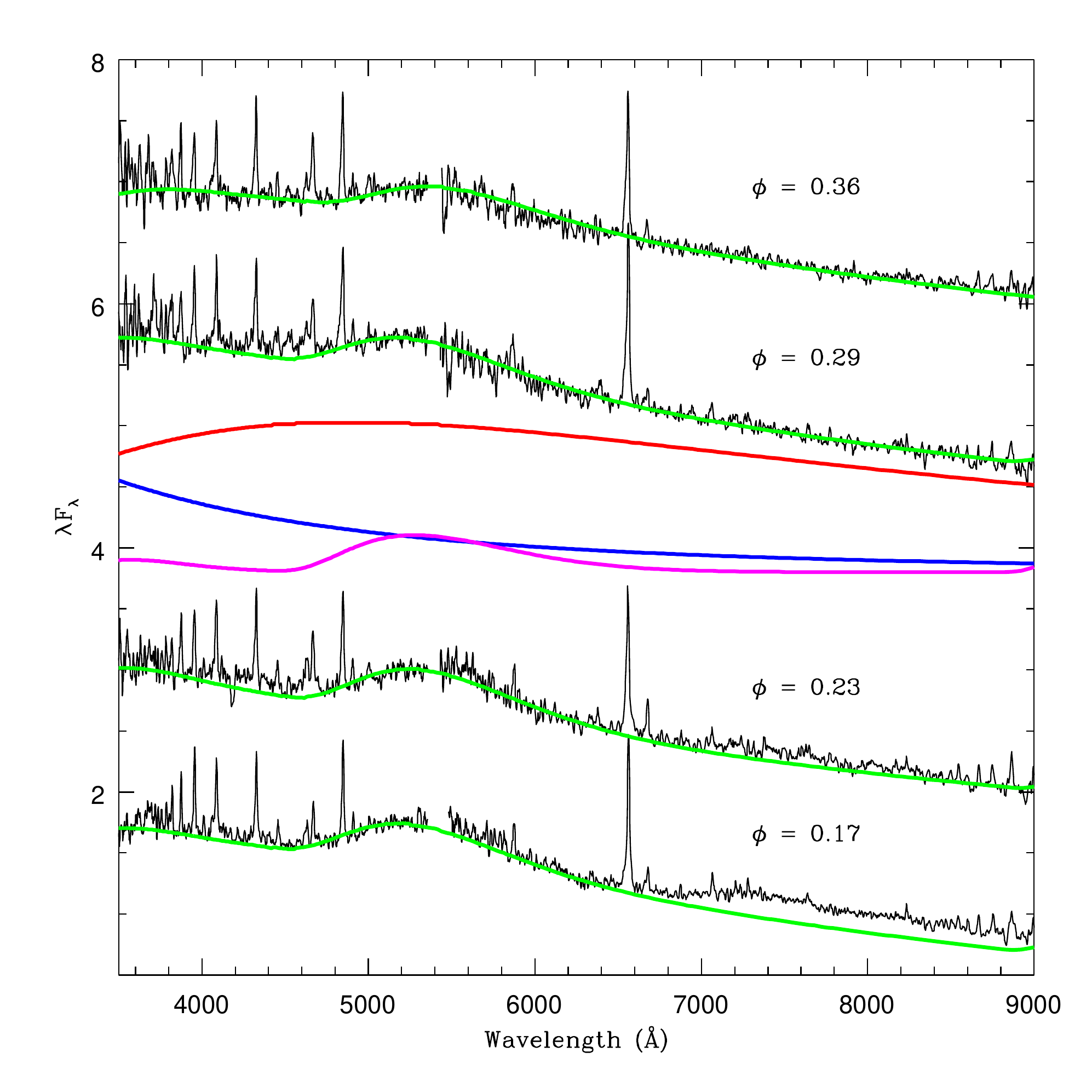}}}
\caption{The spectra (black) for the phase interval 0.17 $\leq$ $\phi$ 
$\leq$ 0.36.  Models (see section 3.2) have been fitted to the data (green). 
For the $\phi$ = 0.29 spectrum, we plot the three spectral components that 
comprise our final model spectrum. This includes a hot blackbody source with
T$_{\rm eff}$ = 54000 K (blue), the irradiated secondary star with 
T$_{\rm eff}$ = 6100 K (red), and a cyclotron spectrum with B = 105 MG, 
$k$T = 16 keV, $\theta$ = 60$^{\circ}$, log$\Lambda$ = 0.0 (magenta). The model
spectra for the other phases in this plot have components with very similar
parameters as those for $\phi$ = 0.29. The spectra in this, and all figures 
that follow, have been boxcar smoothed by 5 pixels. They have also
been vertically offset by arbitrary values for clarity. See Fig. \ref{optlc} 
for approximate $UBVRI$ fluxes.}
\label{fig2a}
\end{figure}
\clearpage

\clearpage
\renewcommand{\thefigure}{2b}
\begin{figure}[htb]
\centerline{{\includegraphics[width=15cm]{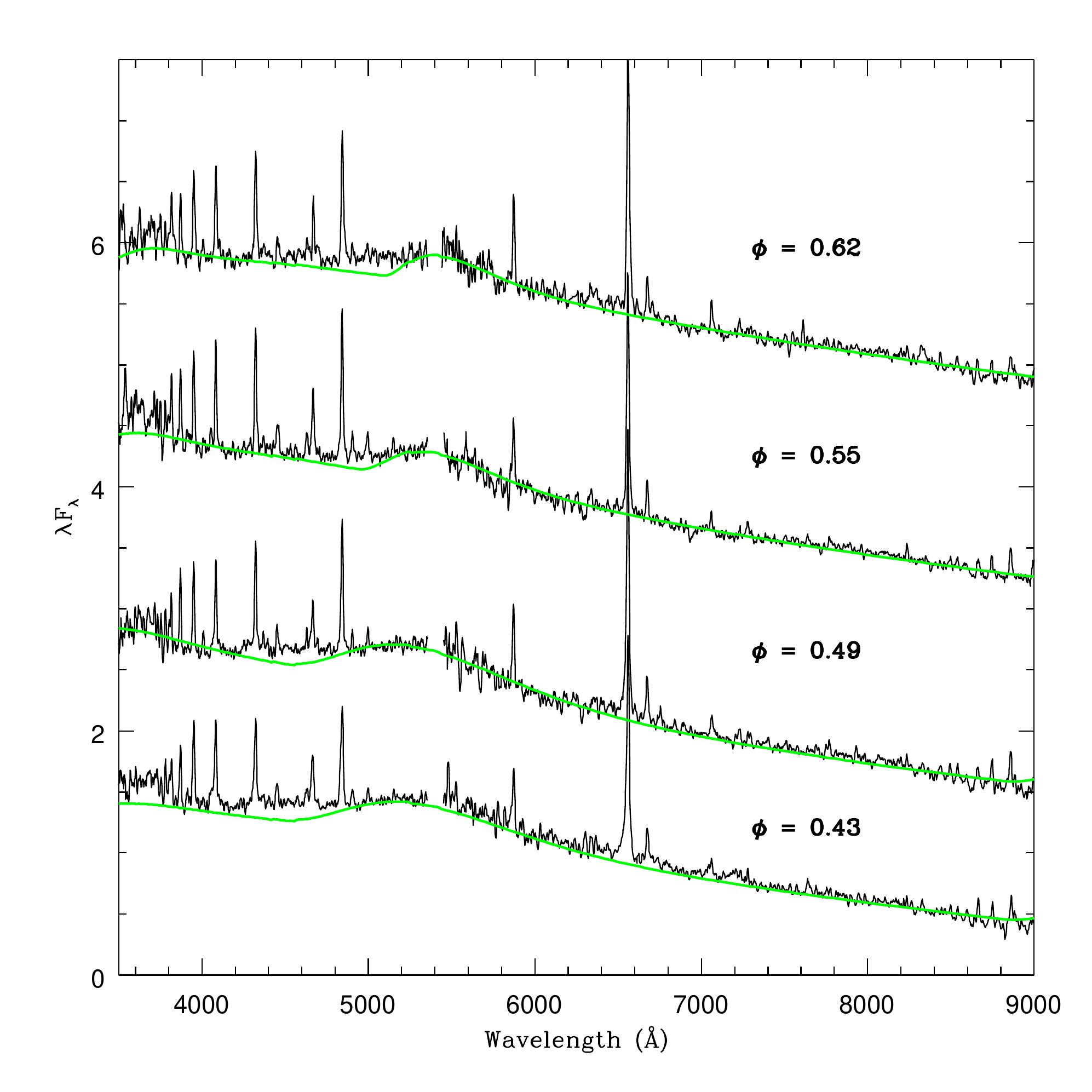}}}
\caption{As in \ref {fig2a}, but for the phase interval 0.43 $\leq$ $\phi$ 
$\leq$ 0.62.}
\label{fig2b}
\end{figure}
\clearpage

\clearpage
\renewcommand{\thefigure}{2c}
\begin{figure}[htb]
\centerline{{\includegraphics[width=15cm]{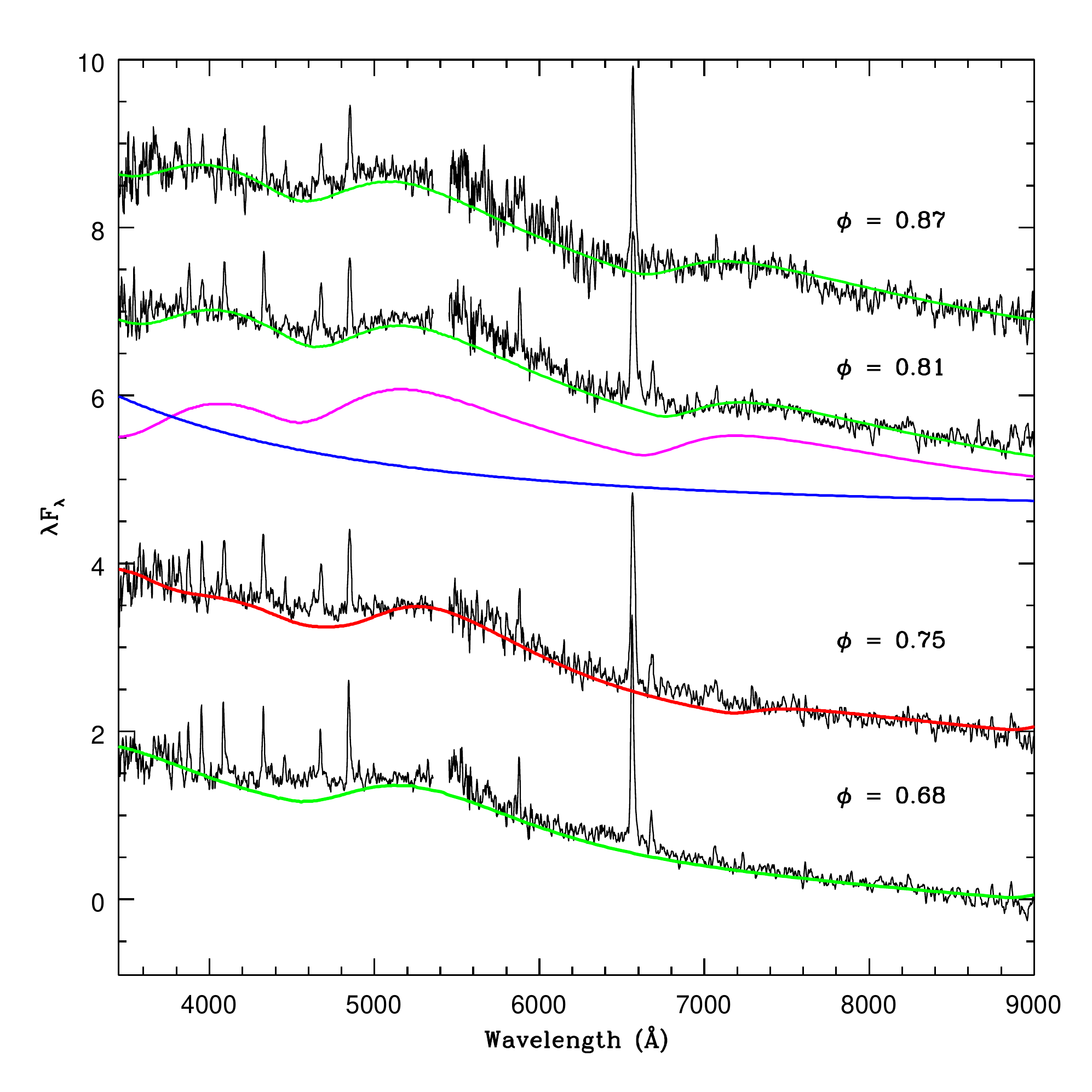}}}
\caption{The spectra for the phase interval 0.68 $\leq$ $\phi$ $\leq$ 0.87. 
The cyclotron model for $\phi$ = 0.68 has B = 105 MG (secondary pole), while 
those for $\phi$ = 0.81 and 0.87 have B = 72 (primary pole). The model for 
$\phi$ = 0.75 (red) results from combining cyclotron models with both of these 
field strengths. The cyclotron emission at this phase is dominated by that 
from the 105 MG pole. We plot the relative contributions of the hot white 
dwarf (blue), and a
cyclotron model with B = 72 MG, $\theta$ = 60$^{\circ}$, $k$T = 35 keV, 
log$\Lambda$ = 1.6 (magenta) for the model (green) of the $\phi$ = 0.81
data set. }
\label{fig2c}
\end{figure}
\clearpage

\renewcommand{\thefigure}{2d}
\begin{figure}[htb]
\centerline{{\includegraphics[width=15cm]{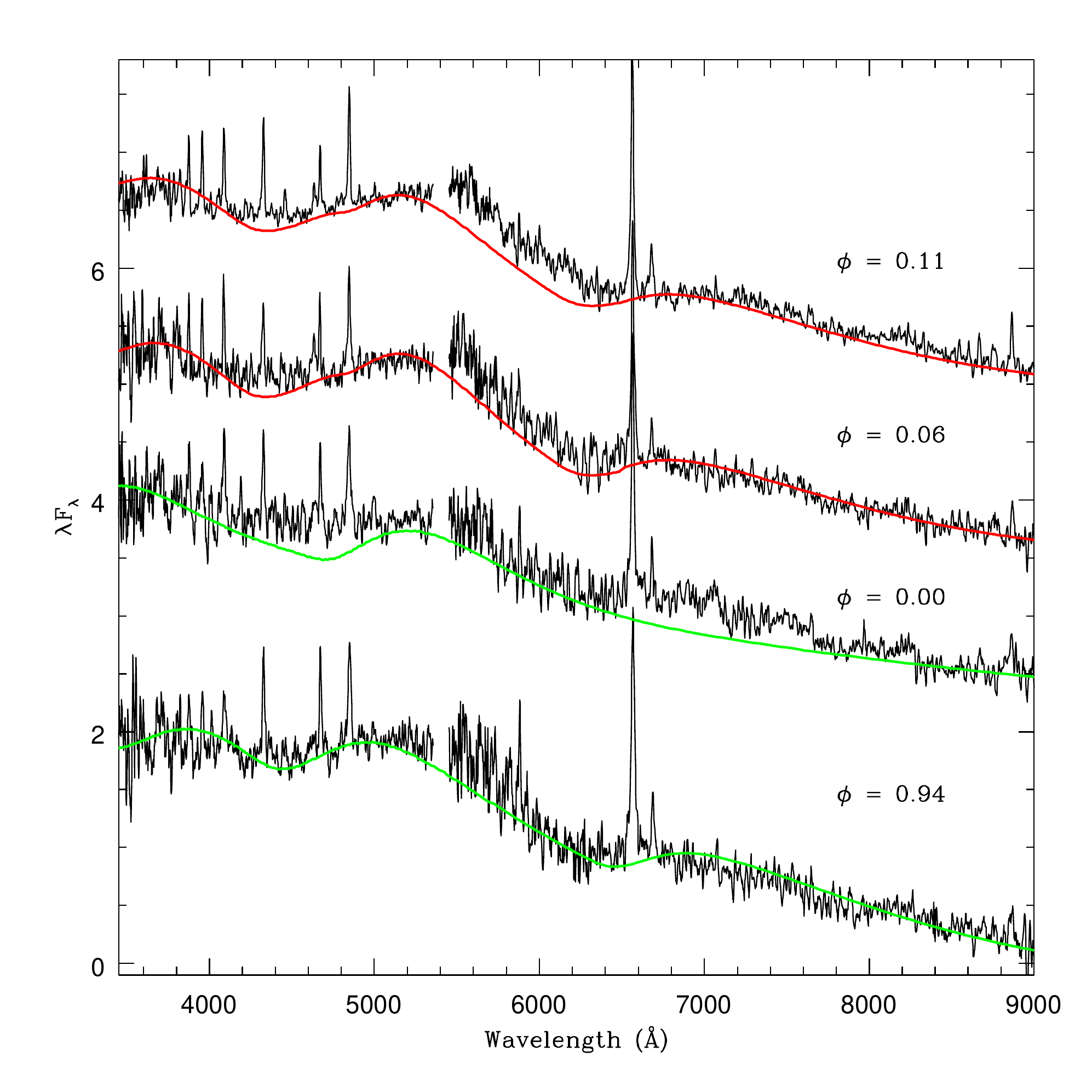}}}
\caption{The spectra of V1500 Cyg covering the phase interval 0.94 $\leq$
$\phi$ $\leq$ 0.11. The green curves are one-component cyclotron models. The
model for $\phi$ = 0.94 has B = 72 MG, while that for $\phi$ = 0 has
B = 105. Cyclotron models that result from a combination of these field
strengths are plotted in red.}
\label{fig2d}
\end{figure}

\clearpage
\renewcommand{\thefigure}{3}
\begin{figure}[htb]
\centerline{{\includegraphics[width=15cm]{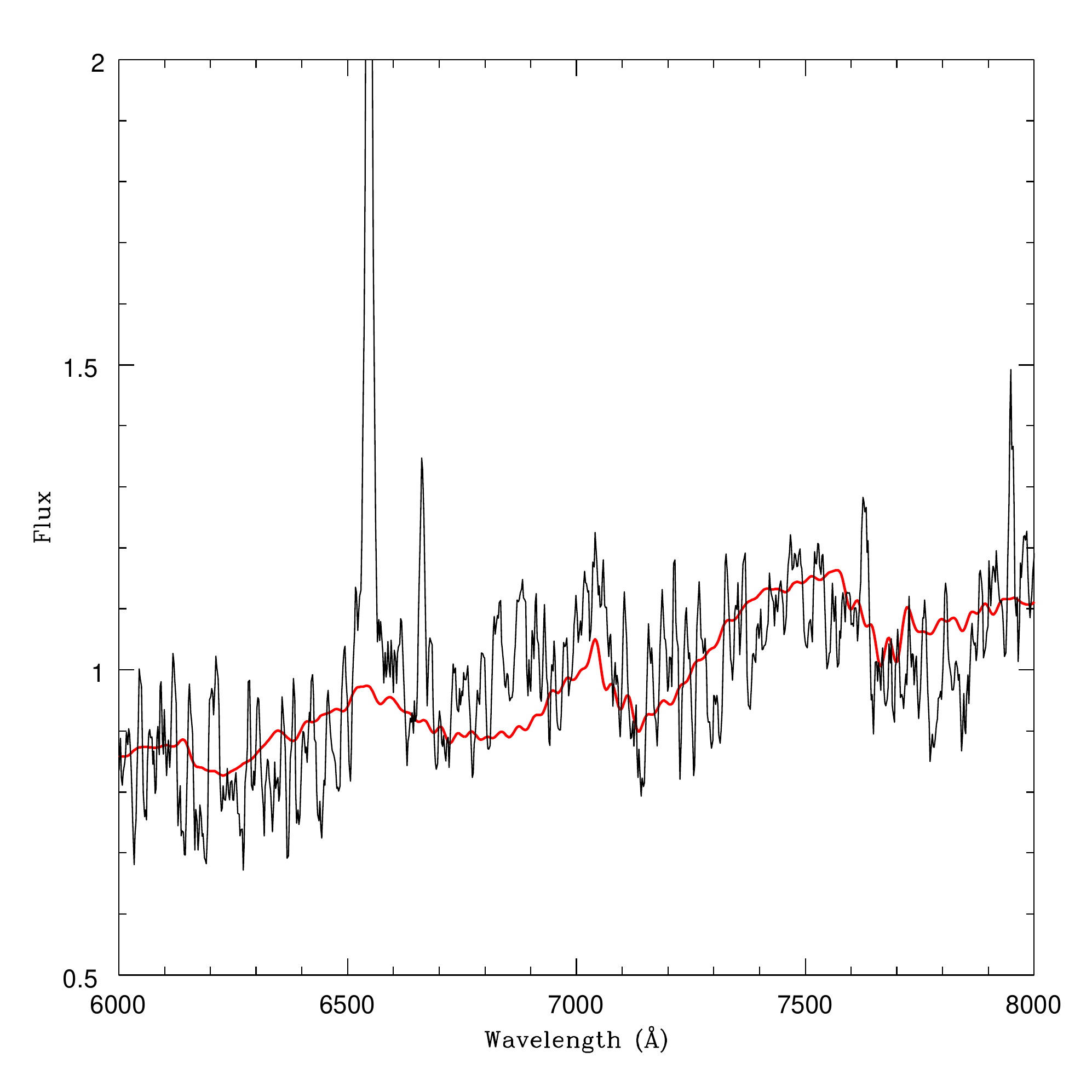}}}
\caption{A comparison of the $\phi$ = 0 spectrum of V1500 Cyg (black) to
that of the M5 dwarf Gl 866A (red).}
\label{M5comp}
\end{figure}
\clearpage

\renewcommand{\thefigure}{4}
\begin{figure}[htb]
\centerline{{\includegraphics[width=13cm]{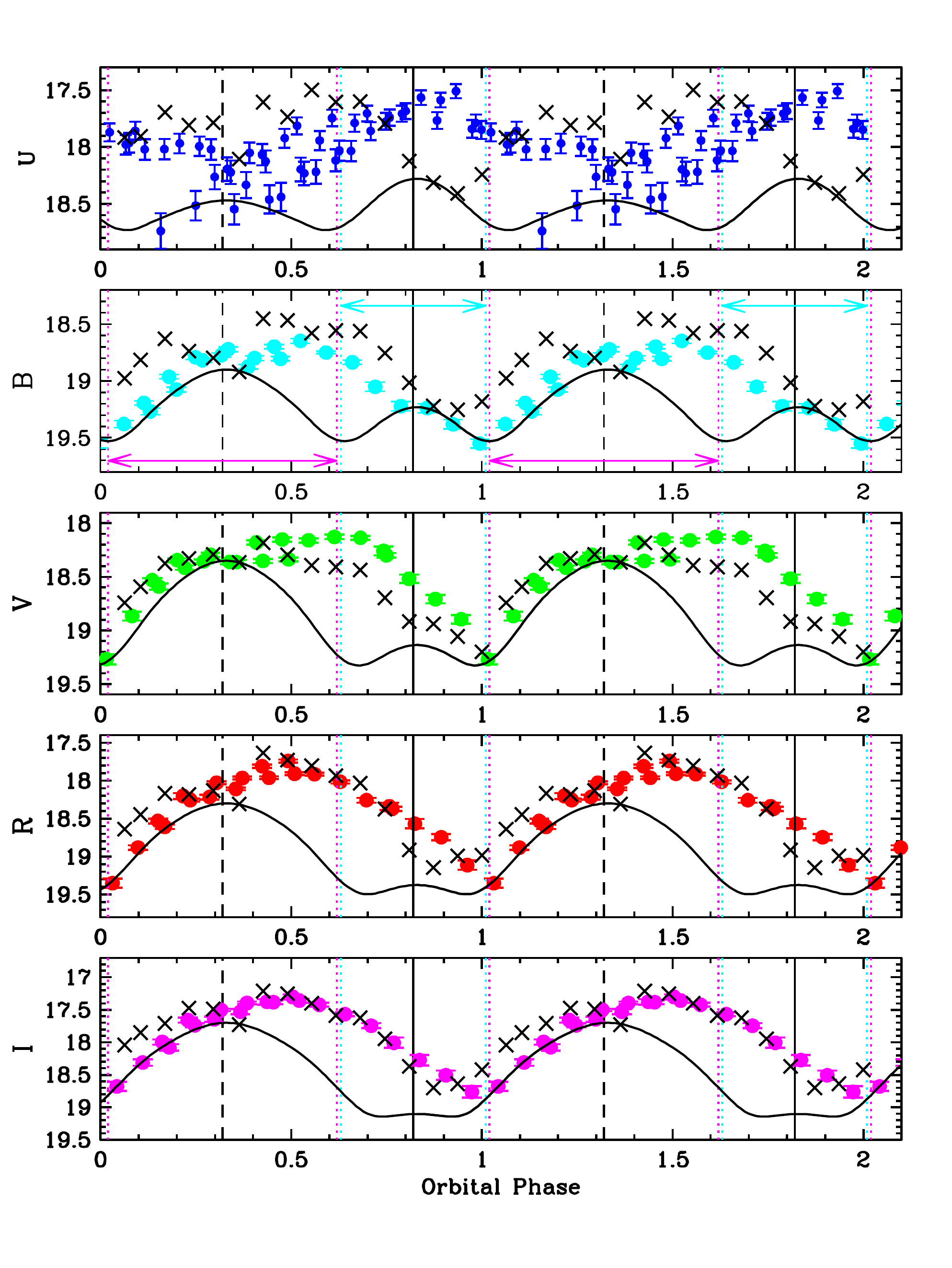}}}
\caption{The $UBVRI$ light curves of V1500 Cyg for 2014 August 31 (from
Harrison \& Campbell 2016). The black crosses represent the photometry
derived from the new spectroscopy. The vertical solid line is the phase of the
transit of the primary X-ray pole, while the dashed line is the transit
of the secondary pole. The dotted cyan lines and arrows (in the $B$-band panel)
indicate the phases that the primary X-ray pole is visible. While the
magenta dotted lines and arrows are for the secondary pole. The solid
black curve is the light curve model described in the text.}
\label{optlc}
\end{figure}

\renewcommand{\thefigure}{5}
\begin{figure}[htb]
\centerline{{\includegraphics[width=15cm]{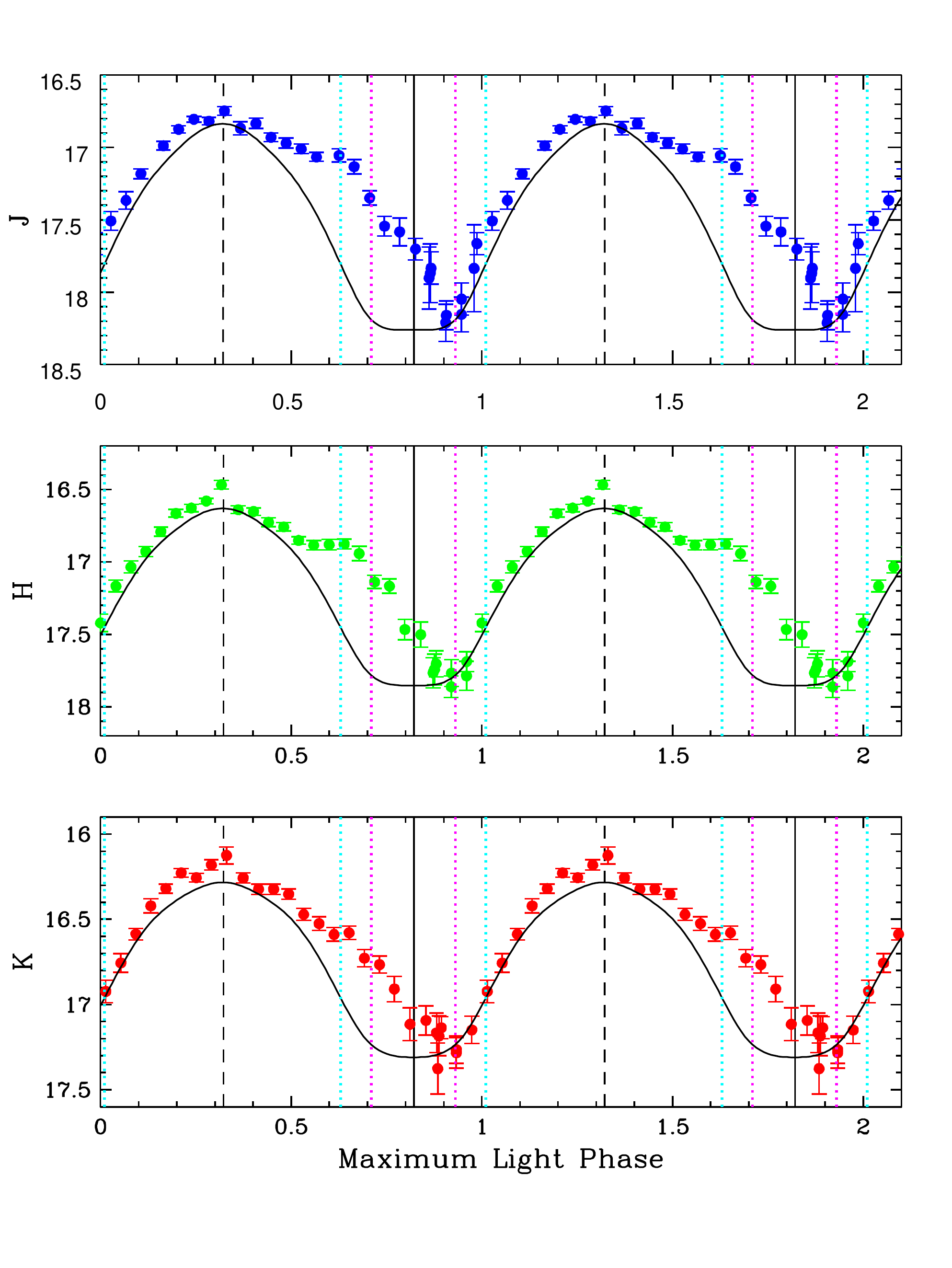}}}
\caption{The $JHK$ light curves of V1500 Cyg for 2014 October 30 (from
Harrison \& Campbell 2016), as in Fig. \ref{optlc}. The dashed, dotted and
solid lines are as those in Fig. \ref{optlc}.}
\label{irlc}
\end{figure}

\end{document}